\RequirePackage{fix-cm}
\documentclass[smallextended]{svjour3}       
\smartqed  
\usepackage{graphicx}
\newcommand{\bea}{\begin{eqnarray}}
\newcommand{\eea}{\end{eqnarray}}
\newcommand{\ben}{\begin{equation}}
\newcommand{\een}{\end{equation}}

\newcommand{\bes}{\begin{subequations}}
\newcommand{\ees}{\end{subequations}}
\usepackage{graphicx,amsmath,times,hhline}
\usepackage{amssymb,graphicx}
\usepackage{amsmath,bm}

\newcommand{\e}{\mbox{e}}
\newcommand{\al}{\alpha}
\newcommand{\ba}{\beta}
\newcommand{\del}{\delta}
\newcommand{\Del}{\Delta}
\newcommand{\ga}{\gamma}

\begin{document}

\title{Degenerate soliton solutions and their dynamics in the nonlocal Manakov system: II Interactions between solitons
}


\author{S. Stalin    \and
        M. Senthilvelan  \and 
	M. Lakshmanan
}


\institute{S. Stalin \at
             Centre for Nonlinear Dynamics, 
	      School of Physics, 
	      Bharathidasan University, 
		Tiruchirappalli-620 024, 
		Tamil Nadu, India
           \and
           M. Senthilvelan (Corresponding Author)\at
              Centre for Nonlinear Dynamics, 
	      School of Physics, 
	      Bharathidasan University, 
		Tamil Nadu, India\\
           \email{velan@cnld.bdu.ac.in}
	\and
           M. Lakshmanan \at
              Centre for Nonlinear Dynamics, 
	      School of Physics, 
	      Bharathidasan University, 
		Tiruchirappalli-620 024, 
		Tamil Nadu, India\\
              \email{lakshman@cnld.bdu.ac.in} 
}

\date{Received: date / Accepted: date}

\maketitle

\begin{abstract}

In this paper, by considering the degenerate two bright soliton solutions of the nonlocal Manakov system,  we bring out three different types of energy sharing collisions for two different parametric conditions.  Among the three, two of them are new which do not exist in the local Manakov equation. By performing an asymptotic analysis to the degenerate two-soliton solution, we explain the changes which occur in the quasi-intensity/quasi-power, phase shift and relative separation distance during the collision process. Remarkably, the intensity redistribution reveals that in the new types of shape changing collisions, the energy difference of soliton in the two modes is not preserved during collision. In contrast to this, in the other shape changing collision, the total energy of soliton in the two modes is conserved during  collision. In addition to this, by tuning the imaginary parts of the wave numbers, we observe localized resonant patterns in both the scenarios. We also demonstrate the existence of bound states in the CNNLS equation during the collision process for certain parametric values.  
\end{abstract}

\section{Introduction}
\label{intro}
Finding new localized wave solutions, studying their dynamics in the nonlocal integrable equations and obtaining new integrable equations from the nonlocal reductions are active areas of research in the study of integrable systems. Recently in \cite{1}, Ablowitz and Musslimani introduced a reverse space nonlocal nonlinear Schr\"{o}dinger (NNLS) equation to explain the wave propagation in a nonlocal medium. A special property associated with this equation is the existence of $\cal{PT}$-symmetry when the self-induced potential obeys the $\cal{PT}$-symmetry condition \cite{2}. The presence of nonlocal field as well as $\cal{PT}$-symmetric complex potential make the nonlocal equations more interesting subject.  Various recent studies have shown that the analysis of NNLS equation and its variant have both physical and mathematical perspectives \cite{3b}-\cite{8a}. Further only a few investigations on the dynamics of solitons in the coupled version of NNLS equation have been reported in the literature \cite{57}-\cite{8a}. In particular, breathing one soliton solution is constructed for the following nonlocal Manakov equation through inverse scattering transform \cite{8a},
\bea
iq_{j,t}(x,t)+q_{j,xx}(x,t)&+&2\sum_{l=1}^{2}q_l(x,t)q_l^{*}(-x,t)q_{j}(x,t)=0, ~j=1,2.
\label{1.1a}
\eea
However the soliton shows singularity in a finite time at $x=0$. The above equation is a vector generalization of reverse space NNLS equation. As we pointed out above Eq. (\ref{1.1a}) also possess self-induced potential $V(x,t)=2\sum_{l=1}^{2}q_l(x,t)q_l^{*}(-x,t)$ and $\cal{PT}$-symmetry since the later obeys the  $\cal{PT}$-symmetric condition. In the first part of accompanying work \cite{8b}, we have constructed general soliton solution for Eq. (\ref{1.1a}) and its augmented version 
\bea
iq_{j,t}^*(-x,t)-q_{j,xx}^*(-x,t)&-&2\sum_{l=1}^{2}q_l^{*}(-x,t)q_l(x,t)q_{j}^{*}(-x,t)=0, ~j=1,2.
\label{1.1b}
\eea
by bilinearizing themt in a non-standard way. The obtained soliton solutions are in general non-singular \cite{8b}. In the present second part we study the dynamics of obtained two-soliton solution which has been reported in the previous part \cite{8b}. Before proceeding further, we first summarize the results presented in \cite{8b}. 

 To construct soliton solution of Eq. (\ref{1.1a}), we also augment Eq. (\ref{1.1b}) given above in the bilinear process. Since, we have treated the fields $q_l(x,t)$ and $q_l^*(-x,t)$, $l=1,2$, present in nonlocal nonlinearity as independent fields, to bilinearize them, we introduce two auxiliary functions, namely $s^{(1)}(-x,t)$ and $s^{(2)}(-x,t)$ in the non-standard bilinear process. By solving the obtained bilinear equations systematically, first we have derived the non-degenerate one soliton solution. From this soliton solution, we have deduced the degenerate one soliton solution and then the solutions which  already exist in the literature. We have also constructed degenerate two-soliton solution for Eq. (\ref{1.1a}). As a continuation of the first part \cite{8b} in the present work we study their dynamics.   

We show that there exists  three types of shape changing collisions in (\ref{1.1a}) for two specific parametric conditions, where the first shape changing collision is similar to the one that arises in the case of local Manakov equation \cite{6a}. 
The second type of shape changing collision is similar to the one that occurs in the mixed CNLS equation \cite{7a2}.  Besides these two collision scenario, we also observe a third type of collision which is a variant of the second type of shape changing collision and it has not been observed in any local 2-CNLS equation. By carrying out the asymptotic analysis for the fields $q_j(x,t)$ and $q_j^*(-x,t)$ in a novel way, we deduce the conservation equation, expression for the phase shift and relative separation distances for all the three collisions.  More surprisingly, in the new types of shape changing collisions, the difference in quasi-intensity of the two modes of a soliton before collision is not equal to the difference in quasi-intensity of the same after collision. However, the total quasi-intensity of solitons before collision is equal to the total quasi-intensity of the solitons after collision in both the modes. In another type of collision the total quasi-intensity of individual solitons as well the total quasi-intensity of soliton before and after collision in both the modes are conserved. Finally, by tuning the imaginary part of the wave numbers we unearth a new type of localized  resonant wave pattern that arises during first and second type of collision processes. We also demonstrate the existence of bright soliton bound states in the nonlocal Manakov equation. 

 The outline of the paper is as follows. In section 2,  by performing an asymptotic analysis, we investigate three types of shape changing collisions via intensity redistribution. Our aim here is to calculate the total energy of the solitons in both the modes, as well as phase shifts and relative separation distances. In Sec. 3, we explain the observation of localized resonant pattern creation during the collision of degenerate two solitons. In this section, we also demonstrate the occurrence of  bound states that occur between the interaction of degenerate two solitons. We present our conclusions in Sec. 4.  

\section{Asymptotic analysis of degenerate two bright nonlocal soliton solution: Shape changing or switching collision}
Differing from the local case,  we  perform the asymptotic analysis on the degenerate two-soliton solutions of Eq. (\ref{1.1a}) and Eq. (\ref{1.1b}).  To perform the asymptotic analysis, we rewrite the two-soliton solution of (\ref{1.1a}) as a nonlinear superposition of two one-solitons. The resultant  expressions are similar to the form given in Eqs. (15a)-(15b) in \cite{8b} but differ in amplitudes and phases. 

 As far as Eqs. (\ref{1.1a}) and (\ref{1.1b}) are concerned, one can identify different types of shape changing collisions. In particular, here we point out three interesting cases, which we designate as Type-I, Type-II and a variant of Type-II collisions. 
\subsection{Type-I shape changing collision}
We visualize Type-I collision for the following choice, namely
\bea
\bar{k}_{1R}<0,~\bar{k}_{2R}>0,~\bar{k}_{1R}<\bar{k}_{2R},~\bar{k}_{1I},~\bar{k}_{2I}>0,~\bar{k}_{1I}<\bar{k}_{2I}\nonumber,\\
k_{1R}>0,~k_{2R}<0,~k_{1R}>k_{2R},~k_{1I},~k_{2I}>0,~k_{1I}<k_{2I} \label{2s.1a},
\eea
where $k_j=k_{jR}+ik_{jI}$, $\bar{k}_j=\bar{k}_{jR}+i\bar{k}_{jI}$, $j=1,2$, For this choice, the nonlocal solitons exhibit a  shape changing collision similar to the one that occurs in the local Manakov equation \cite{6a}. We call this type of collision as local Manakov type collision or Type-I collision. For the parametric restrictions (\ref{2s.1a}) the solitons $S_1$ and  $S_2$ are well separated initially. The variables $\xi_{jR}$ and $\bar{\xi}_{jR}$'s in the two nonlocal solitons behave asymptotically as (i) $\xi_{1R}$, $\bar{\xi}_{1R}\sim 0$, $\xi_{2R}$, $\bar{\xi}_{2R}\rightarrow \pm\infty$ as $t\pm\infty$ (soliton 1 ($S_1$)) and (ii) $\xi_{2R}$, $\bar{\xi}_{2R}\sim 0$, $\xi_{1R}$, $\bar{\xi}_{1R}\rightarrow \mp\infty$ as $t\pm\infty$ (soliton 2 ($S_2$)). Here the variables, $\xi_{jR}$ and $\bar{\xi}_{jR}$ are the real parts of the wave variables $\xi_j$ and $\bar{\xi}_{j}$ and are equal to $-k_{jI}(x+2k_{jR}t)$ and $-\bar{k}_{jI}(x-2\bar{k}_{jR}t)$,  respectively. 
\subsection{Type-II shape changing collision and its variant}
 The nonlocal solitons exhibit another interesting collision scenario for the following parametric condition:  
\bea
\bar{k}_{1R}>0,~\bar{k}_{2R}<0,~\bar{k}_{1R}>\bar{k}_{2R},~\bar{k}_{1I},~\bar{k}_{2I}>0,~\bar{k}_{1I}<\bar{k}_{2I}\nonumber\label{2s.1b},\\
k_{1R}<0,~k_{2R}>0,~k_{1R}<k_{2R},~k_{1I},~k_{2I}>0,~k_{1I}<k_{2I}.
\eea
For this choice, the nonlocal Manakov equation admits a collision similar to the one that occurs in the local mixed CNLS equation \cite{7a2}. To differentiate this second type of collision from the earlier one which is pointed out in the previous paragraph we call this collision as nonlocal mixed CNLS like collision or Type-II collision. For the parametric restriction given in (\ref{2s.1b}) the wave variables $\xi_{jR}$ and $\bar{\xi}_{jR}$'s  behave asymptotically as (i) $\xi_{1R}$, $\bar{\xi}_{1R}\sim 0$, $\xi_{2R}$, $\bar{\xi}_{2R}\rightarrow \mp\infty$ as $t\pm\infty$ (soliton 1 ($S_1$)) and (ii) $\xi_{2R}$, $\bar{\xi}_{2R}\sim 0$, $\xi_{1R}$, $\bar{\xi}_{1R}\rightarrow \pm\infty$ as $t\pm\infty$ (soliton 2 ($S_2$)). 

 For the same parametric restriction chosen in (\ref{2s.1b}) we also come across another type of new shape changing collision which we call as variant of Type-II collision which has not been observed in any local $2$-CNLS equation. Thus in the nonlocal Manakov equation, we observe three types of shape changing collisions whereas in the local Manakov equation we come across only one type of shape changing collision \cite{6a}. We perform the asymptotic analysis for all the three types of shape changing collisions. Our results show that the asymptotic analysis carried out on Type-II and its variant collisions match with each other.  

\subsection{Asymptotic forms in Type-I shape changing collision}

 Now we can check that the parametric choice given in (\ref{2s.1a}) for the Type-I collision leads to the following asymptotic forms. 

\underline{(i) Before Collision:} ($t\rightarrow-\infty$)\\
\hspace{1.5cm} 
In the limit $t\rightarrow-\infty$, the two-soliton solution reduces to the following two independent one soliton solutions:\\

(a) \underline{Soliton 1:} ($\xi_{1R}$, $\bar{\xi}_{1R})\sim 0$, $\xi_{2R}$, $\bar{\xi}_{2R}\rightarrow +\infty$
\bes
\bea
q_j(x,t)=\frac{A_j^{1-}(k_1+\bar{k}_1)\e^{\frac{(\bar{\xi}_{1R}-\xi_{1R})}{2}+i\frac{(\bar{\xi}_{1I}-\xi_{1I})}{2}}}{2i[\cosh(\chi_{jR}^{1-})\cos(\chi_{jI}^{1-})+i\sinh(\chi_{jR}^{1-})\sin(\chi_{jI}^{1-})]},
\eea 
 where $A_j^{1-}=\frac{i}{(k_1+\bar{k}_1)}\e^{\rho_j^1-\frac{\theta_j^1-}{2}}$,~$\rho_j^1=\ln\al_1^{(j)}$,~$\theta_j^{1-}=\Del_1^{(j)}-\rho_j^1$,~
$\chi_{jR}^{1-}=\frac{\xi_{1R}+\bar{\xi}_{jR}+\theta_{jR}^{1-}}{2}$,\\$\chi_{jI}^{1-}=\frac{\xi_{1I}+\bar{\xi}_{1I}+\theta_{jI}^{1-}}{2}$,~$j=1,2$. Here, subscript $j$ ($=1,2$) represents the modes $q_1$ and $q_2$, respectively, and the superscript denotes the soliton 1 at $t\rightarrow-\infty$. The parameters $A_j^{1-}$ and $\theta_{jR}^{1-}$ denote the amplitude and phase of the soliton 1 in both the components before collision. The corresponding asymptotic analysis on the fields $q_j^{*}(-x,t)$ yields the following expression,
\bea
q_j^{*}(-x,t)=\frac{\hat{A}_j^{1-}(k_1+\bar{k}_1)\e^{\frac{-(\bar{\xi}_{1R}-\xi_{1R})}{2}-i\frac{(\bar{\xi}_{1I}-\xi_{1I})}{2}}}{2i[\cosh(\hat{\chi}_{jR}^{1-})\cos(\hat{\chi}_{jI}^{1-})+i\sinh(\hat{\chi}_{jR}^{1-})\sin(\hat{\chi}_{jI}^{1-})]},\label{as}
\eea 
where 
$\hat{A}_j^{1-}=\frac{i}{(k_1+\bar{k}_1)}\e^{\hat{\rho}_j^1-\frac{\hat{\theta}_j^1-}{2}}$,~$\hat{\rho}_j^1=\ln\ba_1^{(j)}$,~$\hat{\theta}_j^{1-}=\ga_1^{(j)}-\hat{\rho}_j^1$,~$\hat{\chi}_{jR}^{1-}=\frac{\xi_{1R}+\bar{\xi}_{jR}+\hat{\theta}_{jR}^{1-}}{2}$,\\$\hat{\chi}_{jI}^{1-}=\frac{\xi_{1I}+\bar{\xi}_{1I}+\hat{\theta}_{jI}^{1-}}{2}$,~$j=1,2$.
In Eq. (\ref{as}), ` hat' corresponds to field $q_j^{*}(-x,t)$.\\
(b) \underline{Soliton 2:} ($\xi_{2R}$, $\bar{\xi}_{2R})\sim 0$, $\xi_{1R}$, $\bar{\xi}_{1R}\rightarrow -\infty$ 
\bea
q_j(x,t)=\frac{A_j^{2-}(k_2+\bar{k}_2)\e^{\frac{(\bar{\xi}_{2R}-\xi_{2R})}{2}+i\frac{(\bar{\xi}_{2I}-\xi_{2I})}{2}}}{2i[\cosh(\chi_{jR}^{2-})\cos(\chi_{jI}^{2-})+i\sinh(\chi_{jR}^{2-})\sin(\chi_{jI}^{2-})]},
\eea
where 
$A_j^{2-}=\frac{i}{(k_2+\bar{k}_2)}\e^{\Del_7^{(j)}-\del_{11}-\frac{\theta_j^{2-}}{2}}$,~$\theta_j^{2-}=\mu_4^{(j)}-\Del_7^{(j)}$,~$\chi_{jR}^{2-}=\frac{\xi_{2R}+\bar{\xi}_{2R}+\theta_{jR}^{2-}}{2}$,~$\chi_{jI}^{2-}=\frac{\xi_{2I}+\bar{\xi}_{2I}+\theta_{jI}^{2-}}{2}$,~$j=1,2$. In the above, amplitude and phase of the soliton 2 before collision is represented by  $A_j^{2-}$ and $\theta_{jR}^{2-}$, respectively. Here, the superscript denotes the soliton 2  before collision. In the same limit, the asymptotic expression of $q_j^{*}(-x,t)$ turns out to be
\bea
q_j^{*}(-x,t)=\frac{\hat{A}_j^{2-}(k_2+\bar{k}_2)\e^{\frac{-(\bar{\xi}_{2R}-\xi_{2R})}{2}-i\frac{(\bar{\xi}_{2I}-\xi_{1I})}{2}}}{2i[\cosh(\hat{\chi}_{jR}^{2-})\cos(\hat{\chi}_{jI}^{2-})+i\sinh(\hat{\chi}_{jR}^{2-})\sin(\hat{\chi}_{jI}^{2-})]},
\eea
where 
$\hat{A}_j^{2-}=\frac{i}{(k_2+\bar{k}_2)}\e^{\ga_7^{(j)}-\del_{11}-\frac{\hat{\theta}_j^{2-}}{2}}$,~$\hat{\theta}_j^{2-}=\varphi_4^{(j)}-\ga_7^{(j)}$,~$\hat{\chi}_{jR}^{2-}=\frac{\xi_{2R}+\bar{\xi}_{2R}+\hat{\theta}_{jR}^{2-}}{2}$,~$\hat{\chi}_{jI}^{2-}=\frac{\xi_{2I}+\bar{\xi}_{2I}+\hat{\theta}_{jI}^{2-}}{2}$,~$j=1,2$.  
\ees\\
\underline{(ii) After Collision:} ($t\rightarrow+\infty$)\\
\hspace{1.5cm} 
In this limit, $t\rightarrow+\infty$, the two-soliton solution reduces to the following two one soliton solutions:\\
 (a) \underline{Soliton 1:} ($\xi_{1R}$, $\bar{\xi}_{1R})\sim 0$, $\xi_{2R}$, $\bar{\xi}_{2R}\rightarrow -\infty$
\bes
\bea
q_j(x,t)=\frac{A_j^{1+}(k_1+\bar{k}_1)\e^{\frac{(\bar{\xi}_{1R}-\xi_{1R})}{2}+i\frac{(\bar{\xi}_{1I}-\xi_{1I})}{2}}}{2i[\cosh(\chi_{jR}^{1+})\cos(\chi_{jI}^{1+})+i\sinh(\chi_{jR}^{1+})\sin(\chi_{jI}^{1+})]},
\eea
where 
$A_j^{1+}=\frac{i}{(k_1+\bar{k}_1)}\e^{\mu_1^{(j)}-\del_{14}-\frac{\theta_j^{1+}}{2}}$,~$\theta_j^{1+}=\mu_5^{(j)}-\mu_1^{(j)}$,~$\chi_{jR}^{1+}=\frac{\xi_{1R}+\bar{\xi}_{1R}+\theta_{jR}^{1+}}{2}$,~$\chi_{jI}^{1+}=\frac{\xi_{1I}+\bar{\xi}_{1I}+\theta_{jI}^{1+}}{2}$,~$j=1,2$. Here, the quantities $A_j^{1+}$ and $\theta_{jR}^{1+}$ define the amplitude and phase of the soliton 1 after collision. In the superscript of the above expressions $1+$ denotes the soliton 1 at $t\rightarrow+\infty$.  
\bea
q_j^*(-x,t)=\frac{\hat{A}_j^{1+}(k_1+\bar{k}_1)\e^{\frac{-(\bar{\xi}_{1R}-\xi_{1R})}{2}-i\frac{(\bar{\xi}_{1I}-\xi_{1I})}{2}}}{2i[\cosh(\hat{\chi}_{jR}^{1+})\cos(\hat{\chi}_{jI}^{1+})+i\sinh(\hat{\chi}_{jR}^{1+})\sin(\hat{\chi}_{jI}^{1+})]},
\eea
where $\hat{A}_j^{1+}=\frac{i}{(k_1+\bar{k}_1)}\e^{\varphi_1^{(j)}-\del_{14}-\frac{\hat{\theta}_j^{1+}}{2}}$,~$\hat{\theta}_j^{1+}=\varphi_5^{(j)}-\varphi_1^{(j)}$,~$\hat{\chi}_{jR}^{1+}=\frac{\xi_{1R}+\bar{\xi}_{1R}+\hat{\theta}_{jR}^{1+}}{2}$,~$\hat{\chi}_{jI}^{1+}=\frac{\xi_{1I}+\bar{\xi}_{1I}+\hat{\theta}_{jI}^{1+}}{2}$,~$j=1,2$.  

(b) \underline{soliton 2:} ($\xi_{2R}$, $\bar{\xi}_{2R})\sim 0$, $\xi_{1R}$, $\bar{\xi}_{1R}\rightarrow -\infty$ 
\bea
q_j(x,t)=\frac{A_j^{2+}(k_2+\bar{k}_2)\e^{\frac{(\bar{\xi}_{2R}-\xi_{2R})}{2}+i\frac{(\bar{\xi}_{2I}-\xi_{2I})}{2}}}{2i[\cosh(\chi_{jR}^{2+})\cos(\chi_{jI}^{2+})+i\sinh(\chi_{jR}^{2+})\sin(\chi_{jI}^{2+})]},
\eea
where $A_j^{2+}=\frac{i}{(k_2+\bar{k}_2)}\e^{\rho_j^2-\frac{\theta_j^{2+}}{2}}$,~$\rho_j^2=\ln\al_2^{(j)}$,~$\theta_j^{2+}=\Del_4^{(j)}-\rho_j^2$, 
$\chi_{jR}^{2+}=\frac{\xi_{2R}+\bar{\xi}_{2R}+\theta_{jR}^{2+}}{2}$,\\$\chi_{jI}^{2+}=\frac{\xi_{2I}+\bar{\xi}_{2I}+\theta_{jI}^{2+}}{2}$,~$j=1,2$. The amplitude and phase of the soliton 2 in the nonlinear Schr\"{o}dinger field after collision is represented by  $A_j^{2+}$ and $\theta_j^{2+}$, respectively. 
\bea
q_j^*(-x,t)=\frac{\hat{A}_j^{2+}(k_2+\bar{k}_2)\e^{\frac{-(\bar{\xi}_{2R}-\xi_{2R})}{2}-i\frac{(\bar{\xi}_{2I}-\xi_{2I})}{2}}}{2i[\cosh(\hat{\chi}_{jR}^{2+})\cos(\hat{\chi}_{jI}^{2+})+i\sinh(\hat{\chi}_{jR}^{2+})\sin(\hat{\chi}_{jI}^{2+})]},
\eea \ees
where $\hat{A}_j^{2+}=\frac{i}{(k_2+\bar{k}_2)}\e^{\hat{\rho}_j^2-\frac{\hat{\theta}_j^{2+}}{2}}$,~$\hat{\rho}_j^2=\ln\ba_2^{(j)}$,~$\hat{\theta}_j^{2+}=\ga_4^{(j)}-\hat{\rho}_j^2$,~ $\hat{\chi}_{jR}^{2+}=\frac{\xi_{2R}+\bar{\xi}_{2R}+\hat{\theta}_{jR}^{2+}}{2}$,\\$\hat{\chi}_{jI}^{2+}=\frac{\xi_{2I}+\bar{\xi}_{2I}+\hat{\theta}_{jI}^{2+}}{2}$,~$j=1,2$. 
One can find the explicit forms of the various constants which appear in the asymptotic forms given in Appendix A.

 Similarly we can calculate the asymptotic forms of Type-II and its variant collisions. However, to avoid too many details, we do not present their explicit forms, but only demonstrate numerically the typical cases.
 
 From the above asymptotic forms of solitons $S_1$ and $S_2$, we conclude that a definite intensity redistribution has occurred among the modes of the nonlocal solitons which can be identified from  the amplitude changes in the solitons $S_1$ and $S_2$. During the collision process, the phases of the solitons have also changed. The  conservation of total energy (or intensity) of the solitons is yet another quantity which characterizes these three shape changing collisions. The conservation of energy which occurs in the Type-II and its variant collisions is entirely different from the collision in the local mixed CNLS equation. In order to   show the intensity redistribution among the modes of the nonlocal solitons from the asymptotic forms,
we calculate the explicit expressions of the amplitudes and phases of the solitons. The obtained expressions of all the quantities which appear from asymptotic forms are given in the Appendix A. 
\subsection{Intensity redistribution}
In this subsection, first we demonstrate how the intensity redistribution and conservation of energy occur between the solitons in the Type-I, Type-II and variant of Type-II collisions.  To demonstrate this, we begin our analysis with the asymptotic forms obtained in the previous sub-section. 
\subsubsection{Intensity redistribution in Type-I collision}
In Type-I collision, the analysis reveals that the amplitudes of the solitons $S_1$ and $S_2$ are changing from $\frac{(k_1+\bar{k}_1)A_j^{1-}}{2i}$ and $\frac{(k_2+\bar{k}_2)A_j^{2-}}{2i}$ to  $\frac{(k_1+\bar{k}_1)A_j^{1+}}{2i}$ and $\frac{(k_2+\bar{k}_2)A_j^{2+}}{2i}$, $j=1,2$, respectively, due to collision.  Similarly the amplitudes of the fields $q_j^{*}(-x,t)$, $j=1,2$ are also changing, during the evolution process, from $\frac{(k_1+\bar{k}_1)\hat{A}_j^{1-}}{2i}$ and $\frac{(k_2+\bar{k}_2)\hat{A}_j^{2-}}{2i}$ to  $\frac{(k_1+\bar{k}_1)\hat{A}_j^{1+}}{2i}$ and $\frac{(k_2+\bar{k}_2)\hat{A}_j^{2+}}{2i}$, $j=1,2$. Here, $A_j^{\pm i}$'s are polarization vectors of the $i$th soliton.  This is because of the energy sharing interaction that occurs between them.  

 In Type-I collision, the quasi-intensity (quasi-power) of the soliton $S_1$ in the first mode $q_1(x,t)$ shares with the soliton $S_1$ in $q_2(x,t)$ mode and the same kind of intensity sharing occurs between the modes of the soliton $S_2$ also. This in turn confirms that the intensity redistribution occurs in between the modes. Even though the intensity redistribution occurs among the solitons that are present in the modes $q_1(x,t)$ and $q_2(x,t)$ the total energy of the individual solitons is conserved  which can be confirmed from     
 
\bes 
\bea
A_1^{1-}\cdot\hat{A}_1^{1-}+A_2^{1-}\cdot\hat{A}_2^{1-}=A_1^{1+}\cdot\hat{A}_1^{1+}+A_2^{1+}\cdot\hat{A}_2^{1+}=1,\\
A_1^{2-}\cdot\hat{A}_1^{2-}+A_2^{2-}\cdot\hat{A}_2^{2-}=A_1^{2+}\cdot\hat{A}_1^{2+}+A_2^{2+}\cdot\hat{A}_2^{2+}=1,
\eea \ees
where the explicit forms of $A_j^{\pm i}$, $i,j=1,2$ are given in Appendix A. In the above, subscripts denote the modes while superscripts represent the soliton number. The above conservation form, reveals the fact that the total quasi-intensity of the individual solitons is conserved. 
In the local case, the total energy of each soliton is calculated  by adding the absolute squares of the amplitudes of the individual modes of the solitons \cite{6a}.  Even though the amplitudes of both the co-propagating solitons are altered after the interaction, the total energy does not vary and is  conserved. 

 In addition to the above, the total energy of the solitons is also conserved. This can be verified by the following conservation form 
\bea
&&A_1^{1-}\cdot\hat{A}_1^{1-}+A_2^{1-}\cdot\hat{A}_2^{1-}+A_1^{2-}\cdot\hat{A}_1^{2-}+A_2^{2-}\cdot\hat{A}_2^{2-}\nonumber\\
&=&A_1^{1+}\cdot\hat{A}_1^{1+}+A_2^{1+}\cdot\hat{A}_2^{1+}+A_1^{2+}\cdot\hat{A}_1^{2+}+A_2^{2+}\cdot\hat{A}_2^{2+}
=2.\label{int1}
\eea
Eq. (\ref{int1}) confirms that the total energy of the solitons $S_1$ and $S_2$ before collision is equal to the total energy of the solitons after collision. 
 
 The change in amplitude of  each one of the solitons in both the components can be evaluated by introducing the transition amplitude $T_j^{l}$ by $T_j^{l}=\frac{A_j^{l+}}{A_j^{l-}}$, where $A_j^{l+}$ is the amplitude of the $l$-th soliton in the $j$-th component after collision and $A_j^{l-}$ is the amplitude of the soliton in the corresponding  mode before collision. To calculate the intensity exchange among the modes of the solitons we multiply the transition amplitude $T_j^l=\frac{A_j^{l+}}{A_j^{l-}}$ by the transition amplitude $\hat{T}_j^{l}=\frac{\hat{A}_j^{l+}}{\hat{A}_j^{l-}}$ of field $q_j^{*}(-x,t)$, where $\hat{A}_j^{l+}$ and $\hat{A}_j^{l-}$ are the amplitudes of the solitons of the field $q_j^{*}(-x,t)$ of each  mode after and before collision respectively. 
This definition also differs from the local Manakov case in which we multiply $T_j^l$ by its own complex conjugate transition element  $T_j^{l*}$, to get  $|T_j^{l}|^2$ \cite{7a}.
\begin{figure}[ht]
\centering
\includegraphics[width=0.3\linewidth]{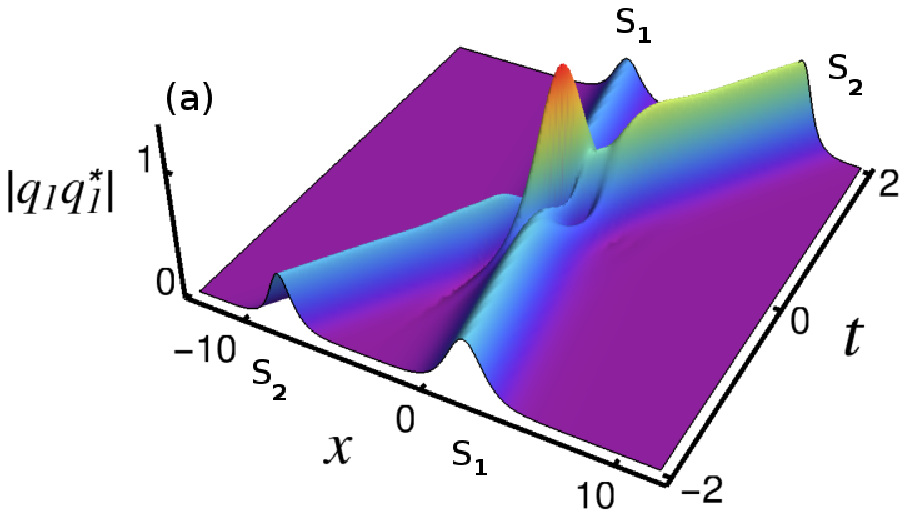}~~~~\includegraphics[width=0.3\linewidth]{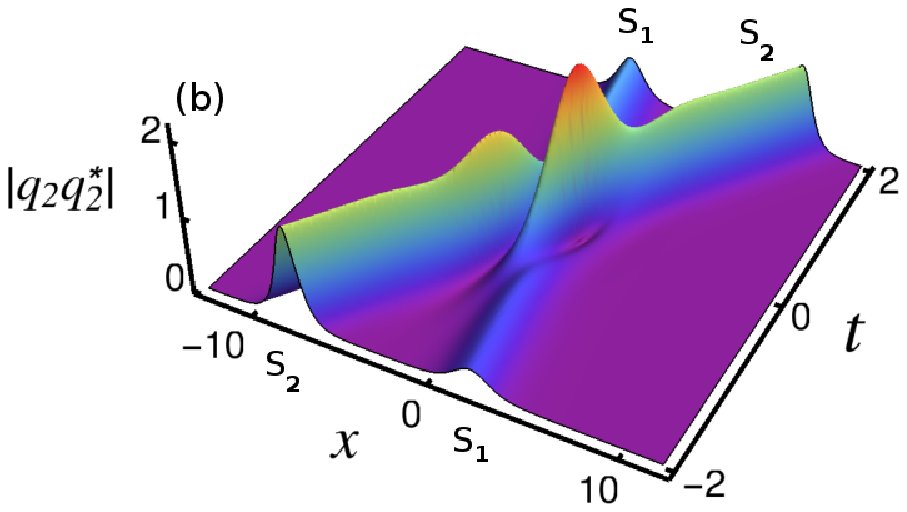}
\caption{Type-I shape changing collision in CNNLS equation: (a) and (b) are the local Manakov type energy sharing collision plotted for the parametric values $k_{1}=0.5+i0.8$, $\bar{k}_1=-0.5+i0.8$, $k_{2}=-2+i$, $\bar{k}_{2}=2+i$, $\al_{1}^{(1)}=1+i$, $\al_{2}^{(1)}=1.5+i$, $\al_{1}^{(2)}=0.5+i$, $\al_2^{(2)}=2+i$, $\ba_{1}^{(1)}=1-i$, $\ba_{2}^{(1)}=-1.5-i$, $\ba_{1}^{(2)}=-0.5-i$ and $\ba_2^{(2)}=2-i$.}
\label{fig2}
\end{figure}
 The intensity exchange between the solitons $S_1$ and $S_2$ due to Type-I collision is defined by
\ben
T_j^{l}\cdot\hat{T}_j^{l}=\frac{A_j^{l+}}{A_j^{l-}}\cdot\frac{\hat{A}_j^{l+}}{\hat{A}_j^{l-}},~~l,j=1,2 \label{i3},
\een
 where all the quantities in the expression  (\ref{i3}) are given in the Appendix A. By suitably fixing the parameters, we can make the right hand side of the above expression to be equal to one. For this special parametric choice, we can come across a pure elastic collision (or shape preserving collision). For all other parametric values there occurs a change in the amplitude in solitons and it leads to the shape changing collision. As in the local CNLS case, one can make one of the transition matrices vanish by suitably fixing the values of the parameters. For this case, the intensity of any one of the solitons in one of the modes becomes zero. 
\subsubsection{Intensity redistribution in Type-II collision and its variant}
 In Type-II collision process also the amplitude of the solitons changes in both the fields $q_j(x,t)$ and $q_j^*(-x,t)$. In this collision scenario, the quasi-intensity of soliton $S_2$ is enhanced in both the modes while the quasi-intensity of soliton $S_1$ is suppressed. This collision scenario is entirely different from the one that occurs in the Type-I collision. A remarkable feature of the Type-II collision is that the total energy of individual solitons is not conserved, so that
\ben
{A}_1^{l-}\cdot\hat{A}_1^{l-}-A_2^{l-}\cdot\hat{A}_2^{l-}\neq A_1^{l+}\cdot \hat{A}_1^{l+}-A_2^{l+}\cdot\hat{A}_2^{l+},~l=1,2.
\label{c1}
\een
From Eq. (\ref{c1}), we infer that the difference in quasi-intensity of soliton $S_1$ in both the modes before collision is not equal to the same  after collision. This is also true for the soliton $S_2$ as well. In the local mixed CNLS equation the energy difference turns out to be the same before and after collision \cite{7a2}.  The free parameters that appear in the degenerate nonlocal two soliton solution (28a)-(28c) given in \cite{8b} do allow the similar kind of shape changing collision as the one happens in the case of local mixed CNLS equation.

 In Type-II shape changing collision, the total intensity of the solitons $S_1$ and $S_2$ in both the components before collision is equal to the the total intensity of the solitons $S_1$ and $S_2$ after collision, that is
\bea
&&A_1^{1-}\cdot\hat{A}_1^{1-}+A_2^{1-}\cdot\hat{A}_2^{1-}+A_1^{2-}\cdot\hat{A}_1^{2-}+A_2^{2-}\cdot\hat{A}_2^{2-}\nonumber\\
&&=A_1^{1+}\cdot\hat{A}_1^{1+}+A_2^{1+}\cdot\hat{A}_2^{1+}+A_1^{2+}\cdot\hat{A}_1^{2+}+A_2^{2+}\cdot\hat{A}_2^{2+}=2.\label{c2}
\eea
The intensity exchange between the solitons in the Type-II collision can also be calculated by defining the transition matrices. In this case the transition matrices are defined by $T_j^{l}\cdot\hat{T}_j^{l}=\frac{A_j^{l+}}{A_j^{l-}}\cdot\frac{\hat{A}_j^{l+}}{\hat{A}_j^{l-}}$, $l,j=1,2$. A special case in which the right hand side becomes one produces shape preserving elastic collision. 
 
In addition to the above Type-II collision, we also observe a variant of it. In the variant of Type-II collision, the intensity of soliton $S_1$ is suppressed in both the modes whereas the intensity of soliton $S_2$ is suppressed in $q_1$ mode and is enhanced in $q_2$ mode. This collision scenario is entirely different from the previous collision processes and has not been encountered  in any local $2$-CNLS equation. The variant of Type-II collision also obeys the non-conservation and conservation relations (\ref{c1}) and (\ref{c2}), respectively.  
 \begin{figure}[ht]
\centering
\includegraphics[width=0.3\linewidth]{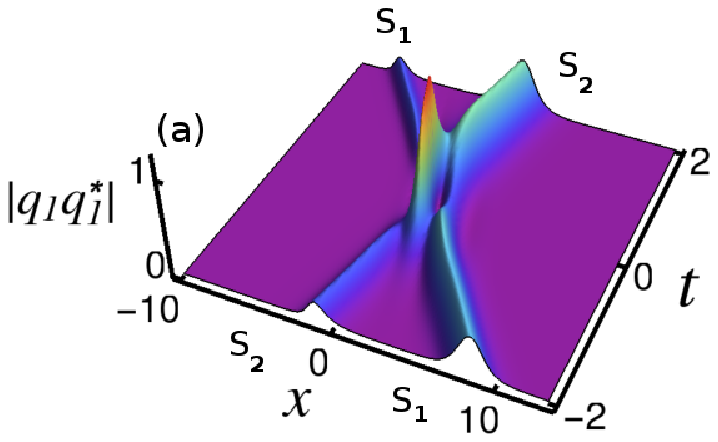}~~~~\includegraphics[width=0.3\linewidth]{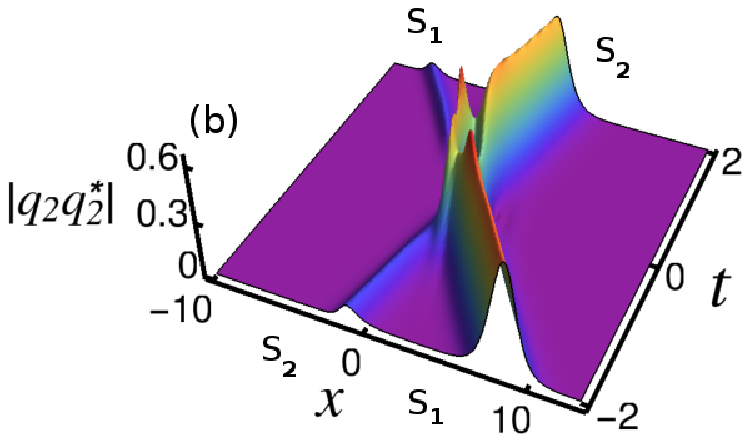}
\caption{Type-II shape changing collision: (a) and (b) are the mixed CNLS like shape changing collision drawn for the parametric values, $k_{1}=-0.5+i0.8$, $\bar{k}_1=0.5+i0.8$, $k_{2}=2+i$, $\bar{k}_{2}=-2+i$, $\al_{1}^{(1)}=1+i$, $\al_{2}^{(1)}=1.5+i$, $\al_{1}^{(2)}=0.5+i$, $\al_2^{(2)}=2+i$, $\ba_{1}^{(1)}=1-i$, $\ba_{2}^{(1)}=-1.5-i$, $\ba_{1}^{(2)}=-0.5-i$, $\ba_2^{(2)}=2-i$.}
\label{fig3}
\end{figure}

 We recall here that  Eq. (\ref{1.1a}) corresponds to  three different equations, namely nonlocal version of (i) Manakov equation, (ii) defocusing CNLS equation and (iii) mixed CNLS equation, depending upon the sign of $\sigma_l$.  It is noted that the shape changing collision that occurs in the local Manakov system differs from the one that occurs in the local mixed CNLS system \cite{7a2}.  For example, the shape changing collision that occurs in the mixed coupled NLS equation can be viewed as an amplification process in which the amplification of signal (say soliton 1) using pump wave (say soliton 2) without any external amplification medium  and without any creation of noise that does not exist in the local Manakov case \cite{6a}. Very surprisingly, the nonlocal Manakov equation simultaneously admits both the types of shape changing collisions  mentioned above, that is the one occurs in the 2-CNLS equation and the other that occurs in the mixed coupled NLS equation. This type of collision has not been observed in any other $(1+1)$-dimensional nonlocal integrable system. 
\begin{figure}[ht]
\centering
\includegraphics[width=0.3\linewidth]{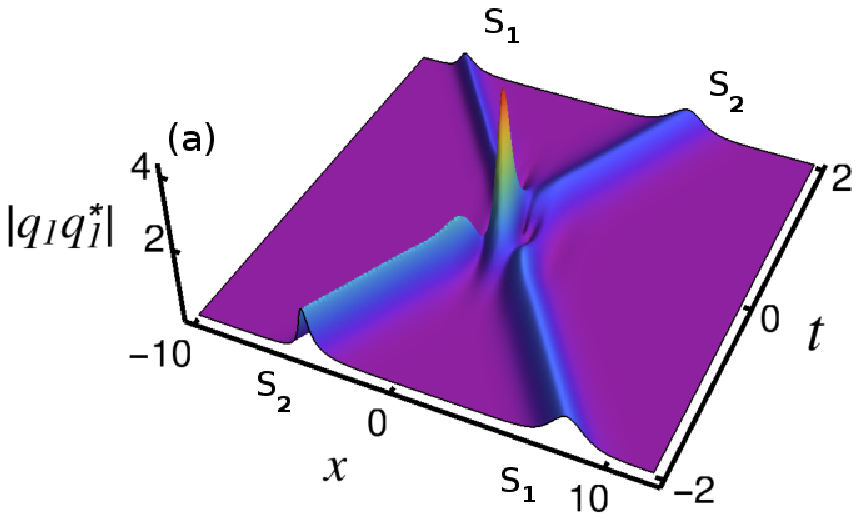}~~~~\includegraphics[width=0.3\linewidth]{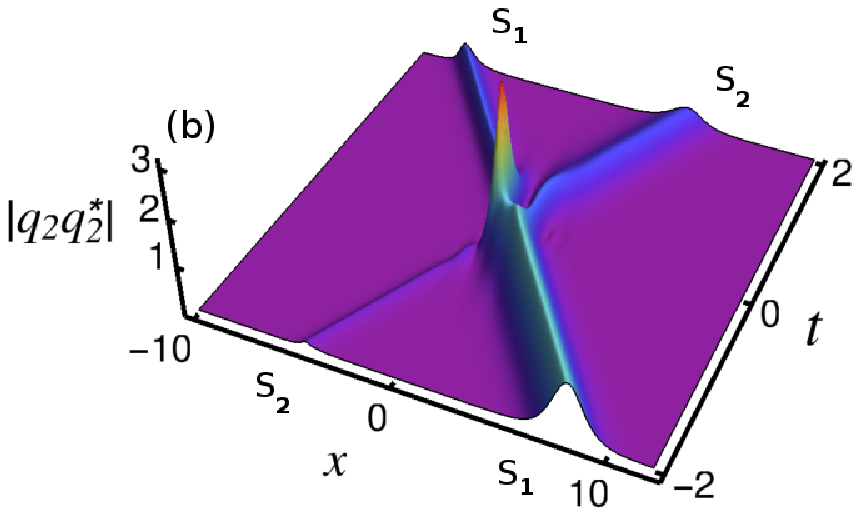}
\caption{A variant of Type-II shape changing collision: (a) and (b) represent the intensity switching collision plotted for the parametric values $k_{1}=-1.5 + i0.8$, $\bar{k}_1=1 +i0.8$, $k_{2}=2+i$, $\bar{k}_{2}=-2+i$, $\al_{1}^{(1)}=1+i$, $\al_2^{(1)}=1.5+i$, $\al_{1}^{(2)}=0.5+i$, $\al_2^{(2)}=2+i$, $\ba_{1}^{(1)}=1-i$, $\ba_{1}^{(2)}=-0.5-i$, $\ba_{2}^{(1)}=-1.5-i$  and $\ba_2^{(2)}=2-i$.}
\label{fig4}
\end{figure}
 In Figs. 2, 3 and 4, we have demonstrated the shape changing collisions that occur in (\ref{1.1a}) for $\sigma_l=+1$. The local Manakov type  shape changing collision that occurs in the system (\ref{1.1a}) is illustrated in Figs. 2a-2b whereas in Figs. 3a-3b  the shape changing collision that occur in (2) as in the case of mixed CNLS equation is shown. A variant of Type-II intensity switching collision is illustrated in Figs. 4a-4b. These three figures also reveal that besides the change in amplitudes, changes also occur in phase shift and relative separation distances. In the following, we calculate these changes.  
\subsection{Phase shifts}
During the collision process, another important quantity, namely the phase is also being altered. The phase identifies essentially the position of the solitons. The change in phase can be calculated from the expressions already obtained. The initial phase of the soliton $S_1$ (=$\frac{\theta_{jR}^{1-}}{2(k_{1I}+\bar{k}_{1I})}$) changes to $\frac{\theta_{jR}^{1+}}{2(k_{1I}+\bar{k}_{1I})}$. Similarly, the initial phase of the soliton $S_2$ (=$\frac{\theta_{jR}^{2-}}{2(k_{2I}+\bar{k}_{2I})}$) changes to $\frac{\theta_{jR}^{2+}}{2(k_{2I}+\bar{k}_{2I})}$. Therefore, the phase shift suffered by the soliton $S_1$ in both the modes during collision is
\bes
\ben
\Phi_1=\frac{1}{2(k_{1I}+\bar{k}_{1I})}\ln\frac{|\rho_{12}\bar{\rho}_{12}(\varrho_1\Gamma_{11}\Gamma_{22}-\varrho_2\nu_1\nu_2-\varrho_3\Gamma_{12}\Gamma_{21})|}{|\Gamma_{11}\Gamma_{22}\kappa_{21}\kappa_{12}|}.
\een
Similarly the phase shift suffered by the soliton $S_2$ is
\ben
\Phi_2=\frac{1}{2(k_{2I}+\bar{k}_{2I})}\ln\frac{|\Gamma_{11}\Gamma_{22}\kappa_{21}\kappa_{12}|}{|\rho_{12}\bar{\rho}_{12}(\varrho_1\Gamma_{11}\Gamma_{22}-\varrho_2\nu_1\nu_2-\varrho_3\Gamma_{12}\Gamma_{21})|}.
\een
From the above two phase shift expressions of solitons $S_1$ and $S_2$, we find that 
\ben
\Phi_2=-\frac{(k_{1I}+\bar{k}_{1I})}{(k_{2I}+\bar{k}_{2I})}\Phi_1 \label{i4}.
\een \ees
 In the Type-II and its variant shape changing collisions, the phase shift suffered by the solitons $S_1$ and $S_2$ is equal to the phase shift suffered by the soliton $S_2$ and $S_1$ in the Type-I collision, respectively. In the Type-II and its variant collisions, the phase shift of the solitons one and two is related by the same relation given in (\ref{i4}). From this relation, we infer that in both the collision processes the soliton $S_2$  gets phase shifted opposite to the soliton  $S_1$. We also find that the phase shifts not only depend on the amplitude parameters $\al_i^{(j)}$ and $\ba_i^{(j)}$, $i,j=1,2$ but also on the wave numbers $k_j$, $\bar{k}_j$.  This is similar to the local Manakov equation and mixed CNLS equation. 
\subsection{Relative separation distances}
The changes which occur in phases of both the solitons in turn cause a change in their relative separation distances during both the collision process. The relative separation distance is nothing but the distance between the positions of the solitons after and before collision \cite{7a}. We denote them by $x_{12}^{\pm}$, where $x_{12}^{+}$ is equal to the position of soliton $S_2$ minus the position of soliton $S_1$ after collision (at $t\rightarrow + \infty$) and $x_{12}^{-}$ is equal to the  position of soliton $S_2$ minus the position of soliton $S_1$ before collision (at $t\rightarrow - \infty$), that is
\ben
x_{12}^{+}=x_2^{+}-x_{1}^{+}, ~~x_{12}^{-}=x_2^{-}-x_{1}^{-},\nonumber
\een
where $x_1^{-}$ and $x_2^{-}$ denote the positions of $S_1$ and $S_2$ at $t\rightarrow -\infty$, respectively, whereas $x_1^{+}$ and $x_2^{+}$ are the positions of $S_1$ and $S_2$ at $t\rightarrow +\infty$, respectively. Their explicit forms can be obtained from the phase shifts of the solitons which turns out to be
\bes
\bea
x_{12}^{-}&=&\frac{1}{2(k_{2I}+\bar{k}_{2I})}\ln\frac{|\rho_{12}\bar{\rho}_{12}(\varrho_1\Gamma_{11}\Gamma_{22}-\varrho_2\nu_1\nu_2-\varrho_3\Gamma_{12}\Gamma_{21})|}{|\Gamma_{11}\kappa_{21}\kappa_{12}\kappa_{22}|}\nonumber\\
&&-\frac{1}{2(k_{1I}+\bar{k}_{1I})}\ln\frac{|\Gamma_{11}|}{|\kappa_{11}|},\\
x_{12}^{+}&=&\frac{1}{2(k_{2I}+\bar{k}_{2I})}\ln\frac{|\Gamma_{22}|}{|\kappa_{22}|}\nonumber\\
&&-\frac{1}{2(k_{1I}+\bar{k}_{1I})}\ln\frac{|\rho_{12}\bar{\rho}_{12}(\varrho_1\Gamma_{11}\Gamma_{22}-\varrho_2\nu_1\nu_2-\varrho_3\Gamma_{12}\Gamma_{21})|}{|\Gamma_{22}\kappa_{21}\kappa_{12}\kappa_{11}|}.
\eea
The total change in relative separation distance is given by
\bea
&&\Del x_{12}=x_{12}^{+}-x_{12}^{-}\nonumber\\
&&\hspace{1.0cm}=\frac{(k_{1I}+\bar{k}_{1I}+k_{2I}+\bar{k}_{2I})}{2(k_{1I}+\bar{k}_{1I})(k_{2I}+\bar{k}_{2I})}\ln\frac{|\Gamma_{11}\Gamma_{22}\kappa_{21}\kappa_{12}|}{|\rho_{12}\bar{\rho}_{12}(\varrho_1\Gamma_{11}\Gamma_{22}-\varrho_2\nu_1\nu_2-\varrho_3\Gamma_{12}\Gamma_{21})|}\nonumber. \eea 
The above expression can also be rewritten as
\bea
&&\Del x_{12}=-\bigg(1+\frac{k_{1I}+\bar{k}_{1I}}{k_{2I}+\bar{k}_{2I}}\bigg)\Phi_1. \label{rs1}
\eea \ees
We observe that the amplitude dependent relative separation distance found above turns out to be the same as in the case of local Manakov equation and mixed CNLS equation. Similarly, we obtain the same expression for the relative separation in all three types of shape changing collisions. It is clear from the  expression (\ref{rs1}), that the relative separation distance non-trivially depends on all the complex parameters, $k_j$, $\bar{k}_j$, $\al_i^{(j)}$'s and  $\ba_i^{(j)}$'s, $i,j=1,2$. Thus the amplitudes, phases and the relative separation distances are all get changed during the interaction between the two nonlocal bright solitons.  
\subsection{Role of complex parameters in the collision process}
From the above results it is clear that all the complex parameters,  $k_j$, $\bar{k}_j$, $\al_i^{(j)}$'s and  $\ba_i^{(j)}$'s, $i,j=1,2$, play important roles in the soliton collision process. In the local  Manakov case, the parameters $\al_i^{(j)}$'s play a crucial role in the shape changing collision process but not the wave numbers \cite{6a}. Hereafter we focus only on the three types of shape changing collisions which occur in the nonlocal Manakov system. 

  In our investigations, we have identified three kinds of collisions. In Type-I collision, the quasi-intensity of soliton $S_2$ is enhanced and the quasi-intensity of soliton  $S_1$ is suppressed in the first mode $q_1(x,t)$. In order to obey the conservation law, the switching of quasi-intensity reversed in the second mode $q_2(x,t)$, that is the quasi-intensity of solitons $S_2$  is suppressed in the first mode whereas the quasi-intensity of soliton $S_1$ is enhanced in the second mode. In this case, the quasi-intensities of solitons are either partially enhanced or partially suppressed. It is demonstrated in Fig. 2a and 2b for the parametric values $k_{1}=0.5+i0.8$, $\bar{k}_1=-0.5+i0.8$, $k_{2}=-2+i$, $\bar{k}_{2}=2+i$, $\al_{1}^{(1)}=1+i$, $\al_{2}^{(1)}=1.5+i$, $\al_{1}^{(2)}=0.5+i$, $\al_2^{(2)}=2+i$, $\ba_{1}^{(1)}=1-i$, $\ba_{2}^{(1)}=-1.5-i$, $\ba_{1}^{(2)}=-0.5-i$ and $\ba_2^{(2)}=2-i$. The second type of shape changing collision is illustrated in Figs. 3a-3b for the parameters $k_{1}=-0.5+i0.8$, $\bar{k}_1=0.5+i0.8$, $k_{2}=2+i$, $\bar{k}_{2}=-2+i$ with all other parameters remaining the same as mentioned above. In these Figs. 3a and 3b, we observe that the intensity of soliton $S_2$ is enhanced in the first mode and a similar change also occurs in the second mode as well. The intensity of soliton $S_1$ is suppressed in both the modes. By comparing the parameter values of Type-I and Type-II collisions, we can easily identify that the only difference in them are signs in the real part of wave number.  All other parameters remain the same. A simple sign change in the real parts of the above parameters causes a dramatic change in the collision dynamics which in turn reveals the strong dependence of this process on complex parameters. The third type of shape changing collision process is demonstrated in Figs. 4a and 4b, for the parametric values $k_{1}=-1.5 + i0.8$, $\bar{k}_1=1 +i0.8$, $k_{2}=2+i$, $\bar{k}_{2}=-2+i$, $\al_{1}^{(1)}=1+i$, $\al_2^{(1)}=1.5+i$, $\al_{1}^{(2)}=0.5+i$, $\al_2^{(2)}=2+i$, $\ba_{1}^{(1)}=1-i$, $\ba_{1}^{(2)}=-0.5-i$, $\ba_{2}^{(1)}=-1.5-i$  and $\ba_2^{(2)}=2-i$. In these figures also we observe that the quasi-intensity of soliton $S_1$ is suppressed in both the modes. In contrast to this the quasi-intensity of soliton $S_2$ is enhanced in $q_2(x,t)$ mode and is suppressed in the $q_1(x,t)$ mode. We note here that all the parameter values are same for Type-II and its variant  collisions except for the values of $k_1$ and $\bar{k}_1$. In all the shape changing collision process the quasi-intensity of solitons in both the components get either enhanced or suppressed. This is because of energy exchange between the modes and the solitons as well. 

 Finally, we note that in the second collision dynamics one may consider the soliton $S_2$ as the signal whereas the soliton $S_1$ as the pump wave (or energy reservoir). In this collision scenario the signals get enhanced or amplified without any use of external amplification medium and without any creation of noise. From these results we conclude that one can use the focusing type nonlocal medium for simultaneously to amplify the signals and to construct the optical computer equivalent to Turing machine in a mathematical sense \cite{7d}. One need not go to separately mixed focusing - defocusing type nonlinear medium for amplifying the signals.   

\begin{figure}[ht]
\centering
\includegraphics[width=0.3\linewidth]{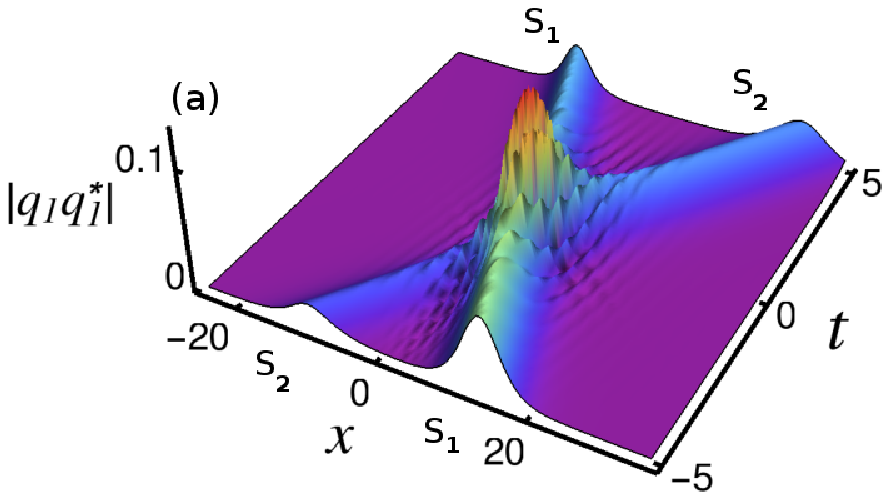}~~~~\includegraphics[width=0.3\linewidth]{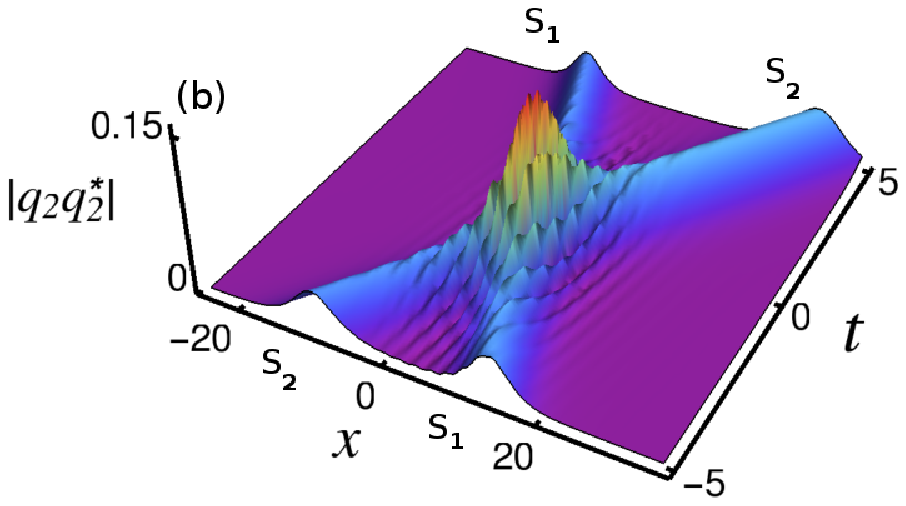}
\caption{(a) and (b) denote the resonant pattern appearing in Type-I collision which is demonstrated for the values $k_{1}=1 + i0.3$, $\bar{k}_1=-1 +i0.3$, $k_{2}=-2+i0.2$, $\bar{k}_{2}=2+i0.2$, $\al_{1}^{(1)}=1+i$, $\al_2^{(1)}=1.5+i$, $\al_{1}^{(2)}=0.5+i$, $\al_2^{(2)}=2+i$, $\ba_{1}^{(1)}=1-i$, $\ba_{1}^{(2)}=-0.5-i$, $\ba_{2}^{(1)}=-1.5-i$  and $\ba_2^{(2)}=2-i$. }
\label{fig5}
\end{figure}

\begin{figure}[ht]
\centering
\includegraphics[width=0.3\linewidth]{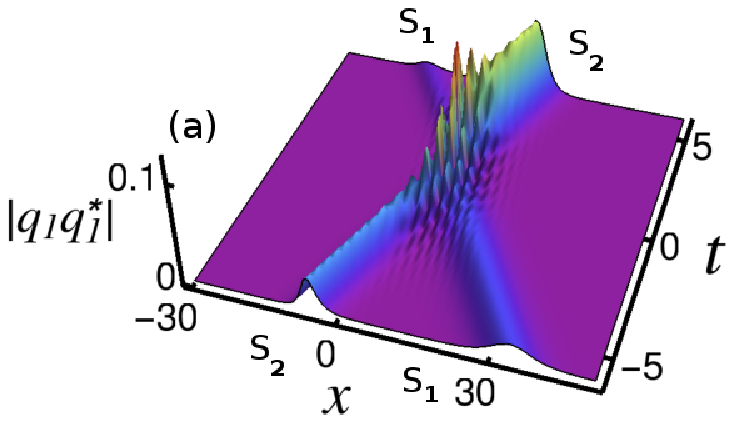}~~~~\includegraphics[width=0.3\linewidth]{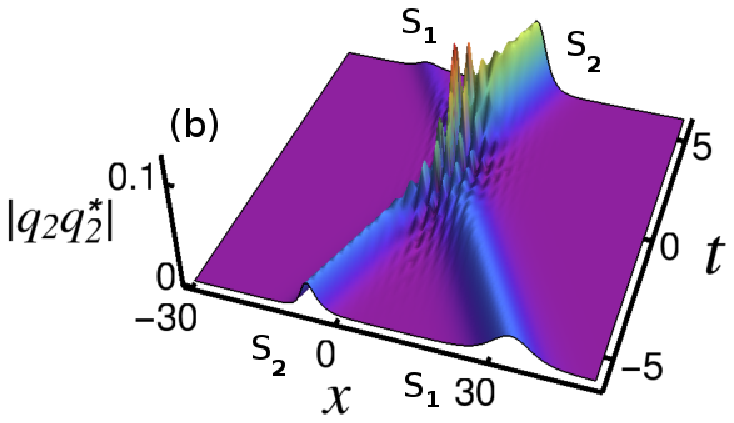}
\caption{(a) and (b) represent the resonant pattern appearing in Type-II collision drawn for the parameter values $k_{1}=-1 + i0.3$, $\bar{k}_1=1 +i0.3$, $k_{2}=2+i0.2$, $\bar{k}_{2}=-2+i0.2$, $\al_{1}^{(1)}=1+i$, $\al_2^{(1)}=1.5+i$, $\al_{1}^{(2)}=1+i$, $\al_2^{(2)}=2+i$, $\ba_{1}^{(1)}=1-i$, $\ba_{1}^{(2)}=-1-i$, $\ba_{2}^{(1)}=1.5-i$  and $\ba_2^{(2)}=2-i$.}
\label{fig6}
\end{figure}
\section{Localized resonant patterns and bright soliton bound states} 
A specific resonant behaviour has been observed during the interaction process in the long wave-short wave resonance interaction system (LSRI system) \cite{r7}. The resonant behaviour occurs exactly in the place at which the phase shift occurs during the collision process. In other words, the localized resonant patterns appear in the interaction regime and it can be considered as an intermediate state. The resonance behaviour was achieved by appropriately choosing the  parameters. One can prolong this intermediate state by fixing the phase shift as large as possible. One can also observe such type of localized resonant behaviour in the present nonlocal Manakov system as well. The resonant behaviour appearing in the Type-I collision is demonstrated in Figs. 5(a)-5(b) for the parametric values $k_{1}=1 + i0.3$, $\bar{k}_1=-1 +i0.3$, $k_{2}=-2+i0.2$, $\bar{k}_{2}=2+i0.2$, $\al_{1}^{(1)}=1+i$, $\al_2^{(1)}=1.5+i$, $\al_{1}^{(2)}=0.5+i$, $\al_2^{(2)}=2+i$, $\ba_{1}^{(1)}=1-i$, $\ba_{1}^{(2)}=-0.5-i$, $\ba_{2}^{(1)}=-1.5-i$  and $\ba_2^{(2)}=2-i$. The same resonant behaviour that one observes in the Type-II shape changing collision can be visualized Figs. 6(a) - 6(b) for the parametric values $k_{1}=-1 + i0.3$, $\bar{k}_1=1 +i0.3$, $k_{2}=2+i0.2$, $\bar{k}_{2}=-2+i0.2$, $\al_{1}^{(1)}=1+i$, $\al_2^{(1)}=1.5+i$, $\al_{1}^{(2)}=1+i$, $\al_2^{(2)}=2+i$, $\ba_{1}^{(1)}=1-i$, $\ba_{1}^{(2)}=-1-i$, $\ba_{2}^{(1)}=1.5-i$  and $\ba_2^{(2)}=2-i$. The resonant behaviours shown in Figs. 5 and 6 are obtained by changing the imaginary values of the wave numbers which we have used to identify the shape changing collision process.  From these figures, we observe that the localized resonant pattern arises in the phase shift regime. In addition to this, we point out that change in the imaginary part of the wave numbers leads to a switching of the Type-I collision into Type-II collision. We also note that the resonant pattern appearing  during the collision process is not same as the one appear in the higher dimensional integrable systems \cite{r7}. In the local Manakov case, one does not observe such behaviour and this occurs only due to the manifestation of nonlocal nature of the system.  We point out that the same type of resonant pattern also appears in the variant of Type-II shape changing collision also.
\begin{figure}[ht]
\centering
\includegraphics[width=0.3\linewidth]{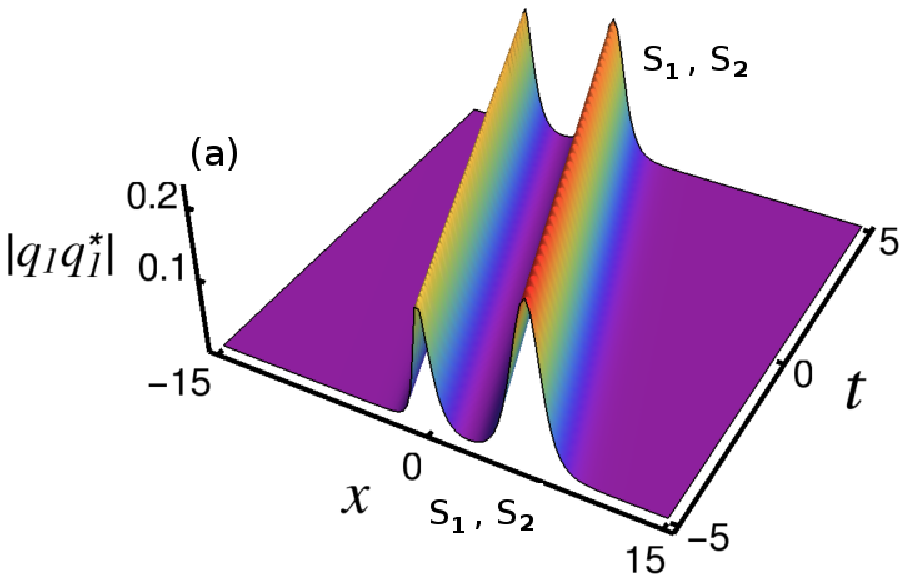}~~~~\includegraphics[width=0.3\linewidth]{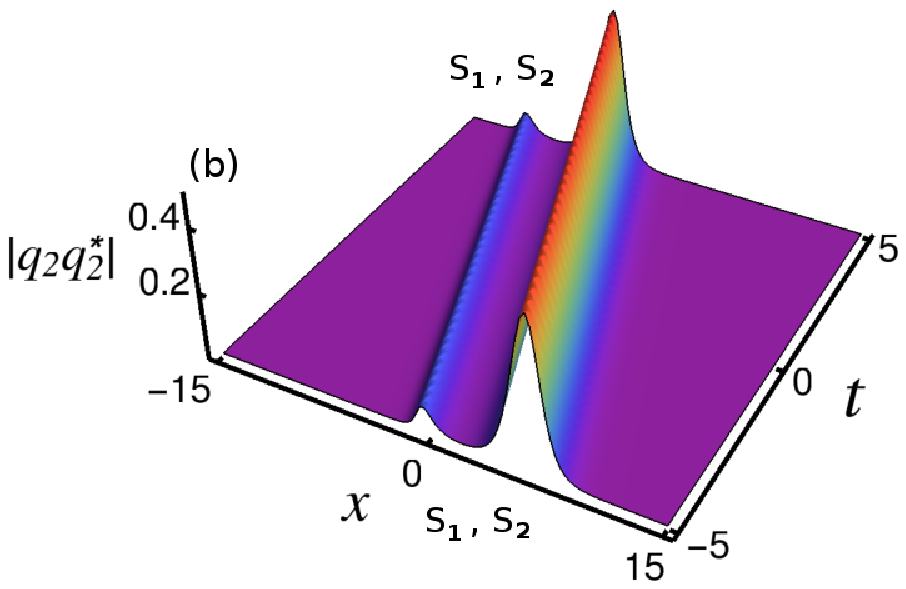}
\caption{(a) and (b) are the parallel propagation of soliton occur in bound state for the parameter values $k_1=0.5+0.8i$ $\bar{k}_1=-0.5+0.8i$, $k_{2}=0.5+0.81i$, $\bar{k}_{2}=-0.5+0.81i$, $\al_{1}^{(1)}=1+i$, $\al_2^{(1)}=1+i$, $\al_{1}^{(2)}=0.1+i$, $\al_2^{(2)}=3+i$, $\ba_{1}^{(1)}=1-i$, $\ba_{1}^{(2)}=-0.1-i$, $\ba_{2}^{(1)}=-1-i$  and $\ba_2^{(2)}=3-i$.}
\label{fig7}
\end{figure}
\begin{figure}[ht]
\centering
\includegraphics[width=0.3\linewidth]{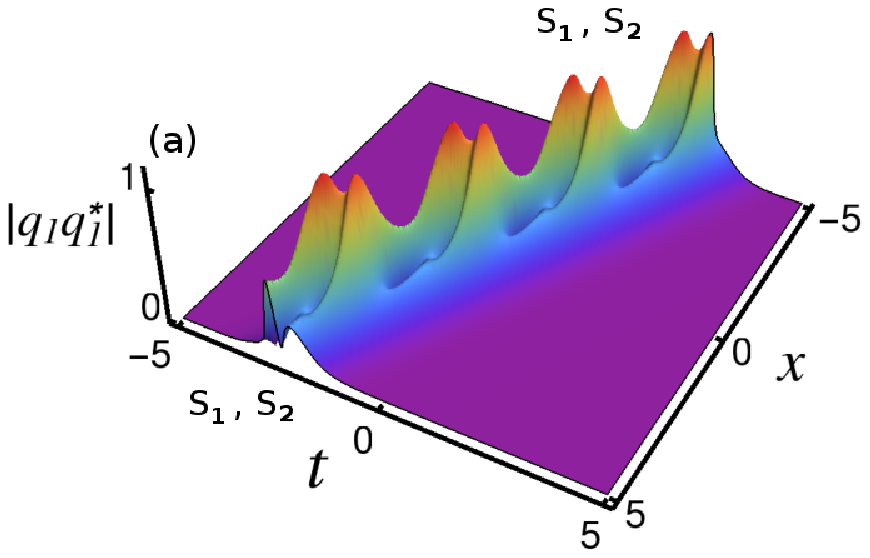}~~~~\includegraphics[width=0.3\linewidth]{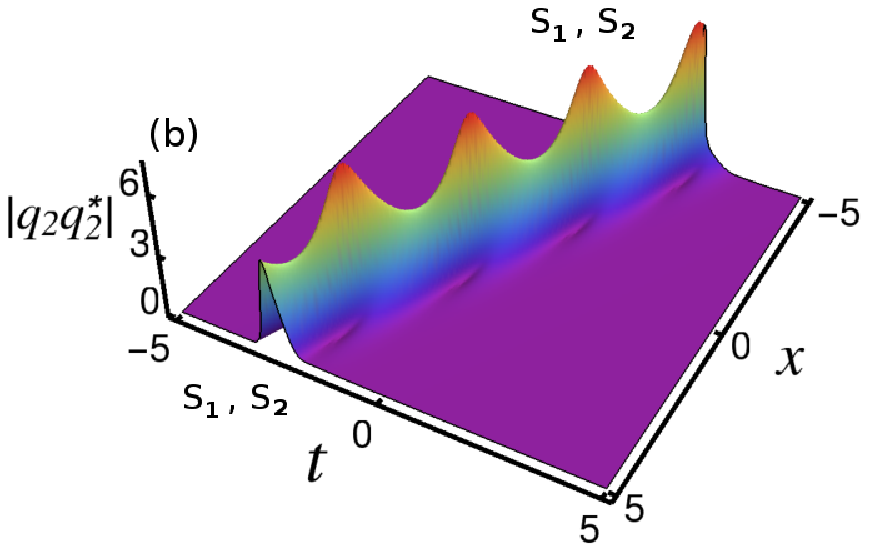}
\caption{(a) denote a novel double hump breathing type bound state occur in the first mode $q_1$ and (b) denotes a single hump breathing type bound state occur in mode $q_2$. The Figs. (a) and (b) drawn for the values $k_1=1+0.8i$ $\bar{k}_1=-1+0.8i$, $k_{2}=1+2.3i$, $\bar{k}_{2}=-1+2.3i$, $\al_{1}^{(1)}=1+i$, $\al_2^{(1)}=1+i0.3$, $\al_{1}^{(2)}=0.3+i$, $\al_2^{(2)}=3+i0.1$, $\ba_{1}^{(1)}=1-i$, $\ba_{1}^{(2)}=-0.3-i$, $\ba_{2}^{(1)}=-1-i0.3$  and $\ba_2^{(2)}=3-i0.1$.}
\label{fig8}
\end{figure}

 Multi-soliton bound states exist in both  integrable and non-integrable systems. When the velocity of the solitons are equal then the solitons bind together to form the bound states \cite{13}. The bound states  can be in different possible forms such as parallel solitons, composite solitons and so on. The parallel soliton bound state exists if the central position of the solitons are different whereas the composite solitons exist when the central positions of the solitons are same \cite{13}. However, these bound state solitons are unstable against small perturbations and they separate into individual solitons which propagate with their identities after some time.  

 In the following, we illustrate the existence of  bound states in the nonlocal Manakov system (\ref{1.1a}).  As we pointed out in section III. B, the soliton velocities are ruled by the parameters $k_{jR}$, $\bar{k}_{jR}$, $j=1,2$, and the central position of the solitons are governed by $\frac{\Del_{1R}}{2(k_{1I}+\bar{k}_{1I})}$ and $\frac{\Del_{2R}}{2(k_{2I}+\bar{k}_{2I})}$ respectively. To explore the parallel soliton bound state, we  fix the parametric values as $k_1=0.5+0.8i$ $\bar{k}_1=-0.5+0.8i$, $k_{2}=0.5+0.81i$, $\bar{k}_{2}=-0.5+0.81i$, $\al_{1}^{(1)}=1+i$, $\al_2^{(1)}=1+i$, $\al_{1}^{(2)}=0.1+i$, $\al_2^{(2)}=3+i$, $\ba_{1}^{(1)}=1-i$, $\ba_{1}^{(2)}=-0.1-i$, $\ba_{2}^{(1)}=-1-i$  and $\ba_2^{(2)}=3-i$. The outcome is displayed in Figs. 7a and 7b. This type of bound state also exhibits oscillatory behaviour as shown in Figs. 8a and 8b for the parameter values $k_1=1+0.8i$ $\bar{k}_1=-1+0.8i$, $k_{2}=1+2.3i$, $\bar{k}_{2}=-1+2.3i$, $\al_{1}^{(1)}=1+i$, $\al_2^{(1)}=1+i0.3$, $\al_{1}^{(2)}=0.3+i$, $\al_2^{(2)}=3+i0.1$, $\ba_{1}^{(1)}=1-i$, $\ba_{1}^{(2)}=-0.3-i$, $\ba_{2}^{(1)}=-1-i0.3$  and $\ba_2^{(2)}=3-i0.1$. As evidenced from this figure one may observe a novel double hump bound soliton state  in the $q_1$ component and a single hump soliton bound state in the $q_2$ component. We point out that the parallel propagation of bound state and the oscillation occurring in the amplitude of the bound state are controlled by the imaginary parts of the wave numbers which appear in the central position of the solitons. We also note here that the oscillations occurring in the amplitude of bound state are usually controlled by the amplitude parameters in any other local integrable systems. 
 
\section{Conclusion}
In this part of the work,  we have brought out the nature of degenerate soliton collisions in the nonlocal Manakov system. In particular, we have brought out  three different types of shape changing collisions for two different parametric conditions. Interestingly  one of them does not exist in the case of local Manakov equation. We have also explained the changes which occur in the quasi-intensity, phase shift and relative separation distance during both the types of energy sharing collisions. We have noticed that in the Type-II and its variant shape changing collisions the difference in the energy of a soliton in the two modes is not preserved during the collision process. However, the total energy of a soliton in the two modes is conserved in the Type-I shape changing collision. We have also demonstrated the occurrence of localized resonant pattern and bound state solutions in the CNNLS equation. Our study gives a better understanding of nonlocal soliton collision in the $\cal{PT}$-symmetric arrays of wave guide systems where the medium exhibits nonlocal nonlinearity. Next we plan to investigate the non-degenerate soliton solutions and their interaction dynamics in some detail.

{\bf \section*{Acknowledgements}}
The work of MS forms part of a research project sponsored by DST-SERB, Government of India, under the Grant No. EMR/2016/001818. The research work of ML is supported by a SERB Distinguished Fellowship and also forms part of the DAE-NBHM research project (2/48 (5)/2015/NBHM (R.P.)/R\&D-II/14127).

{\bf \section*{Appendix}}
\section*{A. Amplitude and phase forms obtained from asymptotic analysis}
The explicit expression for the amplitudes and phases of the solitons 1 and 2 before and after collision ($t\rightarrow\pm\infty$) obtained from the asymptotic analysis of Type-I collision are given below:
The amplitude and phase of the soliton 1 $S_1$ before collision are
\bes\bea 
A_j^{1-}=\frac{\al_1^{(j)}}{\Gamma_{11}^{1/2}},~\hat{A}_j^{1-}=\frac{\ba_1^{(j)}}{\Gamma_{11}^{1/2}},~\theta_{jR}^{1-}=\ln\frac{|\Gamma_{11}|}{|\kappa_{11}|}\label{app1}.
\eea
The amplitude and phase of the soliton 2 $S_2$ before collision are
\bea
&&\hspace{-1cm}A_j^{2-}=\frac{(\kappa_{21}\bar{\varrho}_{12})^{1/2}\bigg((-1)^{j}k_1\ba_1^{(3-j)}\nu_1-\bar{k}_2\al_2^{(j)}\Gamma_{11}+\bar{k}_1\al_1^{(j)}\Gamma_{21}\bigg)}{(\Gamma_{11}\kappa_{12}\varrho_{12})^{1/2}\bigg(\varrho_{1}\Gamma_{11}\Gamma_{22}-\varrho_{2}\nu_1\nu_2-\varrho_{3}\Gamma_{21}\Gamma_{12}\bigg)^{1/2}},\\
&&\hspace{-1cm}\hat{A}_j^{2-}=\frac{(\kappa_{12}\varrho_{12})^{1/2}\bigg(-k_2\ba_2^{(j)}\Gamma_{11}+k_1\ba_1^{(j)}\Gamma_{12}+(-1)^{(3-j)}\bar{k}_1\al_1^{(3-j)}\nu_2\bigg)}{(\Gamma_{11}\kappa_{21}\bar{\varrho}_{12})^{1/2}\bigg(\varrho_{1}\Gamma_{11}\Gamma_{22}-\varrho_{2}\nu_1\nu_2-\varrho_{3}\Gamma_{21}\Gamma_{12}\bigg)^{1/2}},\\
&&\hspace{-1cm}\theta_{jR}^{2-}=\ln\frac{|\bar{\varrho}_{12}\varrho_{12}(\varrho_{1}\Gamma_{11}\Gamma_{22}-\varrho_{2}\nu_1\nu_2-\varrho_{3}\Gamma_{21}\Gamma_{12})|}{|\Gamma_{11}\kappa_{12}\kappa_{21}\kappa_{22}|}\label{app2},
\eea
The amplitude and phase of the soliton 1 $S_1$ after collision are
\bea
&&\hspace{-1cm}A_j^{1+}=\frac{(\kappa_{12}\bar{\varrho}_{12})^{1/2}\bigg((-1)^{j}k_2\ba_2^{(3-j)}\nu_1-\bar{k}_2\al_2^{(j)}\Gamma_{12}+\bar{k}_1\al_1^{(j)}\Gamma_{22}\bigg)}{(\Gamma_{22}\kappa_{21}\varrho_{12})^{1/2}\bigg(\varrho_{1}\Gamma_{11}\Gamma_{22}-\varrho_{2}\nu_1\nu_2-\varrho_{3}\Gamma_{21}\Gamma_{12}\bigg)^{1/2}},\\
&&\hspace{-1cm}\hat{A}_j^{1+}=\frac{(\kappa_{21}\varrho_{12})^{1/2}\bigg(-k_2\ba_2^{(j)}\Gamma_{21}+k_1\ba_1^{(j)}\Gamma_{22}+(-1)^{(3-j)}\bar{k}_2\al_2^{(3-j)}\nu_2\bigg)}{(\Gamma_{22}\kappa_{12}\bar{\varrho}_{12})^{1/2}\bigg(\varrho_{1}\Gamma_{11}\Gamma_{22}-\varrho_{2}\nu_1\nu_2-\varrho_{3}\Gamma_{21}\Gamma_{12}\bigg)^{1/2}},\\
&&\hspace{-1cm}\theta_{jR}^{1+}=\ln\frac{|\bar{\varrho}_{12}\varrho_{12}(\varrho_{1}\Gamma_{11}\Gamma_{22}-\varrho_{2}\nu_1\nu_2-\varrho_{3}\Gamma_{21}\Gamma_{12})|}{|\Gamma_{22}\kappa_{11}\kappa_{21}\kappa_{12}|}\label{app3}.
\eea
The amplitude and phase of the soliton 2 $S_2$ after collision are
\bea
&&A_j^{2+}=\frac{\al_2^{(j)}}{\Gamma_{22}^{1/2}},~\hat{A}_j^{2+}=\frac{\ba_2^{(j)}}{\Gamma_{22}^{1/2}},
~\theta_{jR}^{2+}=\ln\frac{|\Gamma_{22}|}{|\kappa_{22}|}\label{app4}.
\eea
\ees
 To verify the non-conservation and conservation relations (\ref{c1}) and (\ref{c2}) for Type-II shape changing collision and its variant, one has to use the expressions of amplitudes and phases of the solitons before collision given in Eqs. (\ref{app1})-(\ref{app2}) for calculating the quantities $A_j^{l+}$ and  $\hat{A}_j^{l+}$. Similarly to calculate the quantities  $A_j^{l-}$ and  $\hat{A}_j^{l-}$ for shape changing collision and its variant, one has to use the expressions of  amplitudes and phases of the solitons after collision given in Eqs. (\ref{app3})-(\ref{app4}).
{\bf \section*{Conflicts of interest}}
 The authors declare that they have no conflict of interest.

\end{document}